\documentclass[aip,jcp,preprint,dvipdfmx]{revtex4-1}
\usepackage{setspace}
\usepackage{amsmath,amssymb}
\usepackage{mathrsfs}
\usepackage{bm}
\usepackage{epstopdf}
\usepackage{color}
\usepackage{graphicx}
\begin{document}


\title{Open Quantum Dynamics of a Three-Dimensional Rotor Calculated Using a Rotationally Invariant System-Bath Hamiltonian: Linear and Two-Dimensional Rotational Spectra}
\author{Yuki Iwamoto}
\email{iwamoto.y@kuchem.kyoto-u.ac.jp}
\author{Yoshitaka Tanimura}
\email{tanimura.yoshitaka.5w@kyoto-u.jp}
\affiliation{Department of Chemistry, Graduate School of Science, Kyoto University, Kyoto 606-8502, Japan}
\date{\today}

\begin{abstract}
We consider a rotationally invariant system-bath (RISB) model in three-dimensional space that is described by a linear rigid rotor independently coupled to three harmonic-oscillator baths through functions of the rotor's Euler angles. While this model has been developed to study the dielectric relaxation of a dipolar molecule in solvation as a problem of classical Debye relaxation, here we investigate it as a problem of open quantum dynamics. Specifically, the treatment presented here is carried out as an extension of a previous work [J. Chem. Phys, 149, 084110 (2018)], in which we studied a two-dimensional (2D) RISB model, to a three-dimensional (3D) RISB model. As in the 2D case, due to a difference in the energy discretization of the total Hamiltonian, the dynamics described by the 3D RISB model differ significantly from those described by the rotational Caldeira-Leggett (RCL) model. To illustrate the characteristic features of the quantum 3D rotor system described by angular momentum and magnetic quantum numbers, we derive a quantum master equation (QME) and hierarchical equations of motion (HEOM) for the 3D RISB model in the high-temperature case. Using the QME, we compute linear and two-dimensional (2D) rotational spectra defined by the linear and nonlinear response functions of the rotor dipole, respectively. The quantum transitions between the angular momentum states and magnetic states arising from polarized Stark fields as well as the system-bath interactions can be clearly observed in 2D rotational spectroscopy.
\end{abstract}

\maketitle

\section{Introduction}\label{sec:intro}
Rotational motion is as important as translational and vibrational motion because it plays a key role in the thermalization process of a molecular system in a condensed phase.\cite{debye, Barrow, Rothschild, Burstein} With advances in experimental and theoretical methods, in particular, multidimensional vibrational spectroscopies and Brownian dynamics theory, the vibrational motion
of complex molecular systems is now well understood, even in the quantum case.\cite{Mukamel, Ishizaki-2006,TanimuraIshizaki, Weiss08,GrabertPR88} Contrastingly, while rotational motion has been investigated with various experimental methods, for example, techniques based on dielectric absorption,\cite{debye} and infrared (IR), far-IR, rotational-Raman,\cite{Burstein,momose} and Terahertz spectroscopies,\cite{Scherer, Nelson1, Nelson2,Nelson3} theoretical investigations of rotational motion, in particular in the quantum case, are lacking.\cite{Lindenberg,Navez,Gelin-1999} This is because rotational motion, in particular in the quantum case, differs in an important way from vibrational motion: While a particle moves in a confined potential in the vibrational case, the frequency is predominantly determined by the inertial motion of the rotor in the rotational case. Moreover, the symmetry of the rotor system is important, in particular in the quantum case. Specifically, the dynamics of two-dimensional and three-dimensional rotors are different in the quantum case, while they are the same in the classical case. Furthermore, when we take into account environmental effects, the situation becomes even more complicated, because we have to maintain the symmetry not only for the rotor system but also for the heatbath with the system-bath coupling. Thus, while the classical description of a Brownian rotor, whose dynamics are equivalent to Langevin dynamics, is appropriate for describing classical rotational relaxation, the quantum description, which has been studied using the rotational Caldeira-Leggett (RCL) model,\cite{Caldeira-AP-1983} does not exhibit rotational bands.\cite{Suzuki-2001,Suzuki-2002, Suzuki-2003} This is because the quantum coherence of the inertial rotor motion is long-lived even at room temperature, while the RCL system exhibits strong relaxation arising from the symmetry breaking of the total rotational Hamiltonian.

In Ref. \onlinecite{Iwamoto}, we investigated the dynamics of a quantum two-dimensional (2D) rotor system using a rotationally invariant system-bath (RISB) model introduced by Gefen, Ben-Jacob and Caldeira in order to study a current-biased tunnel junction.\cite{Caldeira-1987} This model consists of a 2D rigid rotor independently coupled to a 2D harmonic oscillator bath through sine and cosine functions of the rotor angle. We found that this model is suitable for the description of rotational spectra, because the model is rotationally invariant, and because the equation of motion based on this Hamiltonian reduces to the Langevin equation in the overdamped limit. With this 2D RISB model, we were able to describe the rotational spectrum from the quantum regime to the classical overdamped regime uniformly as a function of the system-bath coupling and bath temperature. 

In the present paper, we extend the previous study to the three-dimensional (3D) case. Such systems have been studied classical mechanically with a Langevin approach as a problem of the dielectric relaxation of a dipolar molecule in solvation, which is known as Debye relaxation.\cite{McConnel,Burstein-1989, Burstein-1992, Filippov-1995, Coffey, Uchiyama1,Stratt1, Stratt2} A rotationally invariant system-bath model in three-dimensional space (3D RISB model) that is described by a linear-rigid rotor independently coupled to three harmonic-oscillator baths through functions of the Euler angles has been employed to derive the Euler-Langevin equation (ELE).\cite{Lindenberg} The same model has been employed to study Debye relaxation in the quantum case on the basis of the quantum master equation.\cite{Navez} Here, we investigate this problem using the framework of open quantum dynamics to analyze complex rotational transitions by computing linear and nonlinear response functions. Because a quantum 3D rotor system is described by the angular momentum and magnetic states, in order to distinguish these states, we must utilize polarized light with an orientation that is different from that of light used for excitation or probing. We investigate transitions through not only the angular momentum but also the magnetic number by computing linear and 2D spectra.\cite{Mukamel,Ishizaki-2006,GallagherJPCA99,GeJPCA02, Hamm,TanimuraIshizaki} 

The organization of this paper is as follows.
In Sec. II, we describe the 3D-RISB model and present the HEOM and QME. The theoretical foundations of linear and 2D spectroscopies are also explained. In Sec. III, we present numerical results and discussion. Section IV is devoted to concluding remarks.

\section{The Model and Theoretical foundation}
\subsection{The 3D-RISB model}
We consider a 3D linear-rigid rotor system described by
\begin{align}
\hat{H}_S &= \frac{\hat{\bf{L}}^2 }{2 I}+ U(\hat{\theta}, \hat{\phi}; t),
\label{eq:RFR}
\end{align}
where $\hat{\bf{L}}$ and $I$ are the angular momentum and the moment of inertia, and $U(\hat{\theta}, \hat{\phi};t)$ is a anisotropic potential that satisfies $ U(\hat{\theta}, \hat{\phi}; t)= U(\hat{\theta}, \hat{\phi}+2\pi; t)$ for $0\le \theta \le \pi$ for the solid angles $\theta$ and $\phi$. {The potential can be time-dependent, for example, for an investigation of laser-induced molecular alignment.\cite{Fleischer2017,Seideman2017}}
The dipole operators of the system are defined as $\hat \mu_{x}=\mu_0 \sin(\hat \theta)\cos(\hat \phi)$, $\hat \mu_{y}=\mu_0 \sin(\hat \theta)\sin(\hat \phi)$, and  $\hat \mu_{z}=\mu_0\cos(\hat \theta)$, where $\mu_0$ is the amplitude of the dipole for the liner-rigid rotor system. The rotor system is independently coupled to three heat baths in the $x$, $y$ and $z$ directions (a three-dimensional heat bath) through the functions of $\theta$ and $\phi$. We can regard these baths to arise from the local electric field due to the surrounding molecules.
The total Hamiltonian is then given by
\begin{align}
\hat{H}_{tot} &= \hat{H}_S + \sum_{\alpha=x,y,z}   \hat{H}_{I+B}^{\alpha},
\label{eq:tot_H}
\end{align}
where  
\begin{align}
\hat{H}_{I+B}^{\alpha} = \sum_k \left\{
\frac{(\hat{p}_{k}^{\alpha})^2}{2 m_k^{\alpha}} + \frac{1}{2}m_k^{\alpha}  (\omega_k^{\alpha})^2 \left(\hat{q}_k^{\alpha}  -
\frac{c_{k}^{\alpha} \hat V_{\alpha}}{m_{k}^{\alpha} (\omega_{k}^{\alpha})^{2}} \right)^2\right\},
\label{eq:Balpha}
\end{align}
and $m_k^{\alpha}$, $\hat{p}_k^{\alpha}$, $\hat q_k^{\alpha}$ and $\omega_k^{\alpha}$ are the mass, momentum, position and frequency variables of the $k$th bath oscillator mode in the $\alpha = x$, $y$ and $z$ direction. 
The system part of the system-bath interactions is defined as $\hat V_{\alpha}\equiv \hat \mu_{\alpha}/\mu_0$, and $c_k^{\alpha}$ is the system-bath coupling constant. 
Here we include the counter terms that are introduced to maintain the translational symmetry of the system Hamiltonian.\cite{TanimuraPRA91} The harmonic baths are characterized by spectral density functions defined as
\begin{align}
J^{\alpha}(\omega) = \frac{\pi}{2} \sum_k \frac{(c_k^{\alpha})^2}{m_k^{\alpha} \omega_k^{\alpha}} \delta(\omega - \omega_k^{\alpha}),
\label{Jomega} 
\end{align}
where $\alpha=x, y$ and $z$. It should be noted that $ J^{\alpha} (\omega)$ need not be the same for different $\alpha$. In particular, they will differ when the surrounding environment is anisotropic. This Hamiltonian without the counter term was introduced by Lindenberg et al. in order to derive the classical Euler-Langevin equation (ELE) for a linear-rigid rotor.\cite{Lindenberg} The ELE for the Ohmic case is presented in Appendix. It is standard to assume the following Drude form for the spectral distribution function in the case of Debye relaxation:
\begin{align}
J^{\alpha}(\omega) = \eta_{\alpha} \frac{{\gamma_{\alpha} ^2 \omega }}
{{\gamma_{\alpha} ^2  + \omega ^2 }}.
\label{JDrude} 
\end{align}
In the Markovian limit, $\gamma \gg \omega_0$, where $\omega_0$ is the characteristic frequency of the system, this reduces to the Ohmic spectral distribution as $ J^{\alpha}(\omega)= \eta_{\alpha} \omega$.

\subsection{HEOM and QME}
The reduced hierarchical equations of motion (HEOM) are reduced equations of motion that can describe the dynamics of the system for non-perturbative and non-Markovian system-bath interactions with any desired accuracy under strong time-dependent perturbations.\cite{Tanimura-1989, Tanimura-1990,Ishizaki-2005,Tanimura-2006,Tanimura-2014,Tanimura-2015}
 In this formalism, the effects of higher-order non-Markovian system-bath interactions are mapped into the hierarchical elements of the reduced density matrix. We can construct the equations of motion for the reduced density operators of the rotor system in similar manner to that of the Brownian system. 

With the extension of the dimension of the hierarchy, this approach is capable of treating 3D heat-bath with the Drude spectral distribution.
Because the quantum nature of the rotor system arises even at high temperature, we do not necessary to include the low temperature correction terms in many realistic situations. The HEOM for the RISB model in the high temperature case is then expressed as\cite{Tanimura-2006}
\begin{eqnarray}
\frac{\partial }
{{\partial t}}\hat \rho _{\{ n_{\alpha} \}} (t) =  - \left( {\frac{\rm i}
{\hbar }\hat H_S^ \times   + \sum_{\alpha=x,y,z}  n_{\alpha}\gamma_{\alpha} } \right)\hat \rho  _{\{ n_{\alpha} \}} (\,t) - \sum_{\alpha=x,y,z}  \frac{\rm i}
{\hbar }\hat V_{\alpha}^ \times  \hat \rho _{\{ n_{\alpha}+1 \}} (t) \nonumber \\
- \sum_{\alpha=x,y,z}  \frac{{{\rm i}n_{\alpha}}}
{\hbar }\hat \Theta_{\alpha} \hat \rho _{\{ n_{\alpha}-1 \}} (t),	
\label{eq:GMmaster}
\end{eqnarray}
where $\{ n_{\alpha} \}\equiv (n_x, n_y, n_z)$  is a set of integers to describe the hierarchy elements and $\{ n_{\alpha}\pm1 \}$ represents, for example, $ (n_x, n_y\pm 1, n_z)$ for $\alpha=y$, and
\begin{eqnarray}
\hat \Theta_{\alpha}  \equiv \eta_{\alpha} \gamma_{\alpha}
\left(\frac{ 1}{\beta}{\hat V_{\alpha}^ \times   - \frac{ \hbar}{2 }\hat H_S^ \times \hat V_{\alpha}^\circ  } \right),
\label{eq:Theta_S}
\end{eqnarray}
with $\hat A^\times \hat \rho \equiv \hat A \hat \rho - \hat \rho \hat A$ and $\hat A^\circ \hat \rho \equiv \hat A \hat \rho + \hat \rho \hat A$ for any operator $\hat A$. We set $\hat \rho _{\{n_{\alpha}-1\}} (t)=0$ for $n_{\alpha}=0$. 
  Note that $\hat H_S^ \times $ in Eq. \eqref{eq:Theta_S} arises because we utilized the integration by part to eliminate the counter term in order to obtain the present HEOM, as in the case of the quantum Fokker-Planck equation.\cite{Tanimura-2015, TanimuraPRA91,TanimuraJCP92} This form is more convenient to take the classical limit, as shown in Ref. \onlinecite{Iwamoto}.  
For $(N_{\alpha} + 1) \gamma_{\alpha} \gg \eta_{\alpha} /\beta$ and $(N_{\alpha} + 1) \gamma_{\alpha} \gg \omega_0$, where $\omega_0=\hbar/2I$ is the characteristic frequency of the system, we can set $ {\rm i}\hat V_{\alpha}^ \times \hat \rho _{\{ N_{\alpha}+1 \}} (t)/\hbar = \hat \Gamma_{\alpha} \hat \rho _{\{ N_{\alpha} \}} (t) $ to truncate the hierarchy, where 
\begin{eqnarray}
\hat \Gamma_{\alpha} \equiv  \frac{{ 1 }}{{\gamma_{\alpha} \hbar^2 }}\hat V_{\alpha}^ \times  \hat \Theta_{\alpha}
\label{eq:Gamma}
\end{eqnarray}
is the damping operator.\cite{Tanimura-2006} For $\gamma_{\alpha}  \gg \omega _0 $ with a fixed value of $\eta_{\alpha} /\gamma_{\alpha}$ for all $\alpha$, we can set $N_{\alpha} =0$ to reduce the HEOM to the QME as
\begin{eqnarray}
\frac{\partial }{{\partial t}}\hat \rho _0 (t) = 
 - {\frac{\rm i}{\hbar }\hat H_S^ \times  } \hat \rho _0 (t) -\sum_{\alpha=x,y,z}  \hat \Gamma_{\alpha} {\rm{\hat \rho }}_0 (t).
\label{eq:Qmaster}
\end{eqnarray}
The explicit form of the damping operator is written as
\begin{align}
\hat \Gamma_{\alpha} {\rm{\hat \rho }}(t)  &=   \frac{\eta_{\alpha}}{\beta \hbar^2}  \left(  [\hat{V}_{\alpha},  \hat{V}_{\alpha} \hat{\rho}(t) ] - [ \hat{V}_{\alpha} ,   \hat{\rho}(t ) \hat{V}_{\alpha}] \right)  \notag \\
& + \frac{\eta_{\alpha}}{2 \hbar^2} \left([\hat{V}_{\alpha} ,  \hat{H}_S \hat{V}_{\alpha} \hat{\rho}(t ) ] + [ \hat{V}_{\alpha} ,   \hat{H}_S \hat{\rho}(t ) \hat{V}_{\alpha} ] - [\hat{V}_{\alpha} ,  \hat{V}_{\alpha} \hat{\rho}(t ) \hat{H}_S] - [ \hat{V}_{\alpha} ,   \hat{\rho}(t ) \hat{V}_{\alpha} \hat{H}_S]  \right) .
\label{eq:QMEiso}
\end{align}
Although we assumed the high temperature heat-bath to derive the HEOM and QME, this condition is easily satisfied for measurements in molecular rotational spectroscopy experiments. For example, for the rotational motion of a HCl, the moment of inertia is $I=2.6 \times 10^{-47} \mathrm{kg \cdot m^2}$, and we have $\beta \hbar \omega_0 \sim 0.05 \ll 1$ at room temperature. {If necessary, we can lower the bath temperature in the framework of the HEOM formalism by including the low-temperature correction terms.\cite{Tanimura-1989, Tanimura-1990,Ishizaki-2005,Tanimura-2006,Tanimura-2014,Tanimura-2015,Lipeng2019}}

\subsection{Linear and 2D rotational spectra}
The HEOM and QME approaches used in this work can be applied to systems with potentials of arbitrary form for the calculation of linear and nonlinear spectra. In the present study, we computed linear absorption (1D) and two-dimensional (2D) spectra for the rotor system. The first-order and third-order response functions are expressed as \cite{Tanimura-2006}
\begin{align}
  R^{(1)}_{\alpha \alpha'} (t)&=\left(\frac{i}{\hbar }\right)\mathrm{Tr}\left\{\hat{\mu}_{\alpha'}\hat{\mathcal{G}}(t)\hat{\mu }_{\alpha}^{\times }\hat{\rho}_{\mathrm{eq}}\right\},
  \label{eq:R1}
\end{align}
and
\begin{align}
  R^{(3)}_{\alpha''' \alpha'' \alpha' \alpha }(t_{3}, t_{2}, t_{1})&=\left(\frac{i}{\hbar }\right)^{3}\left\{\hat{\mu}_{\alpha'''}\hat{\mathcal{G}}(t_{3})\hat{\mu }_{\alpha''}^{\times }\hat{\mathcal{G}}(t_{2})\hat{\mu}_{\alpha'}^{\times }\hat{\mathcal{G}}(t_{1})\hat{\mu}_{\alpha}^{\times }\hat{\rho}_{\mathrm{eq}}\right\},
  \label{eq:R3}
\end{align}
where $\hat{\mathcal{G}}(t)$ is Green's function in the absence of a laser interaction evaluated from Eq.\ \eqref{eq:GMmaster} or Eq.\eqref{eq:Qmaster}, and $\hat{\rho }_{\mathrm{eq}}$ is the equilibrium state.

To evaluate these response functions, we developed a computational
program incorporating the HEOM or QME presented in Eq.\ \eqref{eq:GMmaster} or Eq.\eqref{eq:Qmaster}.  We first ran the
computational program to evaluate Eq.\ \eqref{eq:R1} or \eqref{eq:R3}
for a sufficiently long time to obtain a true thermal equilibrium
state, $\hat{\rho }_{\mathrm{eq}}$.  The system
was excited by the first interaction, $\hat{\mu }_{\alpha}^{\times }$, at $t=0$. The evolution of the perturbed elements were then computed by running the
program for the HEOM or QME up to some time $t_{1}$. The linear response
function, $R_{\alpha \alpha'}^{(1)}(t)$, defined in Eq.\ \eqref{eq:R1}, was then calculated from the expectation value of $\hat{\mu}_{\alpha'}$, while the third-order response function, $R_{\alpha''' \alpha'' \alpha' \alpha }^{(3)}(t_{3}, t_{2}, t_{1})$, defined in Eq.\ \eqref{eq:R3}, were calculated from the expectation value of $\hat{\mu}_{\alpha'''}$ after applying the operators $\hat{\mu }_{\alpha'}^{\times }$ and $\hat{\mu }_{\alpha''}^{\times }$ by running the program to the period $t_{2}$ and $t_{3}$, respectively.
The rotational absorption spectrum
and the 2D correlation spectrum are evaluated as\cite{GallagherJPCA99,GeJPCA02, Ishizaki-2006,TanimuraIshizaki,Hamm}
\begin{align} 
  I_{\alpha' \alpha }^{\mathrm{(abs)}}(\omega _{1})&= \mathrm{Im}\int _{0}^{\infty }\mathrm{d}t_{1}e^{i\omega _{1}t_{1}}R_{\alpha' \alpha }^{(1)}(t_{1})
  \label{eq:I1}
\end{align}
and
\begin{align} 
  I_{\alpha''' \alpha'' \alpha' \alpha }^{\mathrm{(corr)}}(\omega _{3},t_{2},\omega _{1})
  &= I_{\alpha''' \alpha'' \alpha' \alpha }^{\mathrm{(NR)}}(\omega _{3},t_{2},\omega _{1}) + I_{\alpha''' \alpha'' \alpha' \alpha }^{\mathrm{(R)}}(\omega _{3},t_{2},\omega _{1}),
  \label{eq:I3}
\end{align}
where the non-rephasing and rephasing
parts of the signal are defined by
\begin{align} 
  I_{\alpha''' \alpha'' \alpha' \alpha }^{\mathrm{(NR)}}(\omega _{3},t_{2},\omega _{1}) =
  \mathrm{Im}\int _{0}^{\infty }\mathrm{d}t_{3}\int _{0}^{\infty }\mathrm{d}t_{1}e^{i\omega _{3}t_{3}}e^{i\omega _{1}t_{1}}R_{\alpha''' \alpha'' \alpha' \alpha }^{(3)}(t_{3}, t_{2}, t_{1}),
  \label{eq:nonreph}
\end{align}
and
\begin{align} 
  I_{\alpha''' \alpha'' \alpha' \alpha }^{\mathrm{(R)}}(\omega _{3},t_{2},\omega _{1}) =
  \mathrm{Im}\int _{0}^{\infty }\mathrm{d}t_{3}\int _{0}^{\infty }\mathrm{d}t_{1}e^{i\omega_{3}t_{3}}e^{-i\omega _{1}t_{1}}R_{\alpha''' \alpha'' \alpha' \alpha }^{(3)}(t_{1}, t_{2}, t_{3}).
  \label{eq:reph}
\end{align}

\section{Results and Discussion}
While we can study dynamics of the 3D rotor system for various physical conditions using HEOM formalism, calculating nonlinear signals is computationally very expensive. Here, we restrict our analysis in the Markovian QME case to explore a characteristic feature of the 3D rotor system by means of 2D rotational spectroscopies.

\subsection{Angular momentum and magnetic quantum numbers}
We represent the eigenstate of the rotor system as $| j ,m \rangle$, where the integers $j \ge 0$ and $|m|\le j$ are the angular momentum quantum number and magnetic quantum number. The dipole operators, $\hat \mu_{\alpha}$ for $\alpha =x, y$, and $z$, are expressed in terms of the operator form of the spherical harmonics $\hat Y^m_j$ as
\begin{align}
\hat \mu_z \propto {{\hat Y}_1^0},~ 
\hat \mu_x \propto ({\hat Y}_1^1 - {\hat Y}_1^{-1}),~{\rm and}~~
\hat \mu_y \propto ({\hat Y}_1^1 + {\hat Y}_1^{-1}).
\end{align}
The dipole operator, $\hat \mu_z$, converts the angular momentum state $| j ,m \rangle$ into $| j \pm 1 ,m \rangle$, whereas $\hat \mu_x$ and $\hat \mu_y$ converts both angular momentum and magnetic quantum state  $| j ,m \rangle$ into $| j \pm 1 ,m \pm 1 \rangle$. \cite{Zare} 
The state representation of the QME for the angular momentum and magnetic quantum number is presented in Appendix B. In the following, we study dynamics of the free rotor in the isotropic environment ($\eta_{x}=\eta_y=\eta_z=\eta$).

\subsection{Effects of polarized Stark fields}
\begin{figure}
\centering
\includegraphics[width=0.99\textwidth]{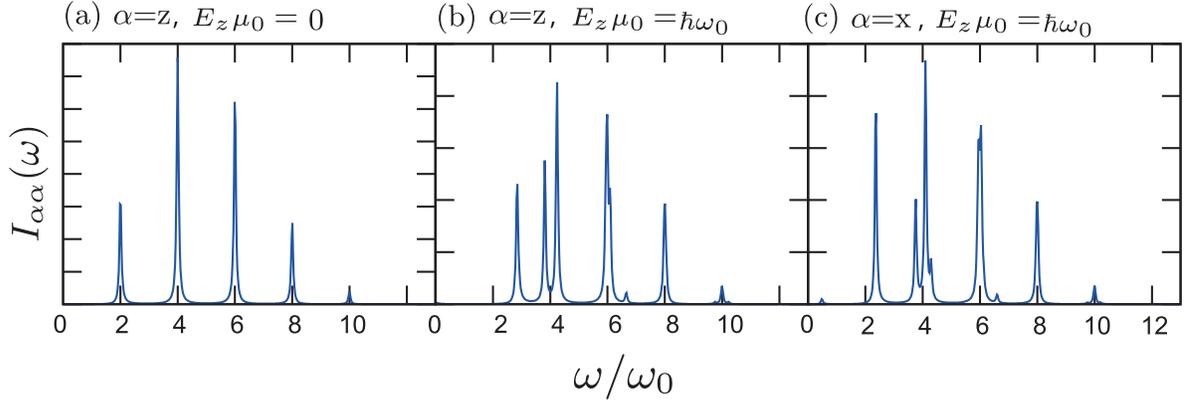}
\caption{
Rotational absorption, $I_{\alpha \alpha}^{\mathrm{(abs)}}(\omega)$, for the (a) $\alpha =z$ polarized light without Stark field, the (b) $\alpha =z$ and (c) $\alpha =x$ polarized light under the $z$ polarized Stark field. Here, we chose the coupling strength, $\eta / \hbar = 0.005$, and the inverse temperature, $\beta\hbar\omega_0=0.2$. The intensity of each line is normalized with respect to its maximum peak intensity.} \end{figure}

\subsubsection{Rotational absorption spectrum}

In the beginning, we explore a characteristic feature of the 3D rotor system utilizing a Stark external field. For this purpose, here we restrict our analysis in the Markovian case with a very weak damping described by the QME given by Eq.\eqref{eq:Qmaster}.
We chose the system-bath coupling $\eta / \hbar = 0.005$ and the inverse temperature, $\beta\hbar\omega_0=0.2$. At this temperature, the angular states $| j, m \rangle$ for $j \leq 6$ with different $|m| \le j$ are thermally well excited by the 3D heat-bath. Here, we consider the $z$ polarized Stark field described by $U(\hat{\theta}, \hat{\phi})=E_{z}\hat \mu_z$, where $E_z$ is the strength of the Stark field. To carry out the numerical simulation, we set ${E_z \mu_0}/{\hbar \omega_0}=1$.

In Fig. 1(a), we plot rotational absorption spectra without the Stark field. In this isotropic case, the spectrum does not depend on the direction of the detection, i.e. $I_{zz}^{\mathrm{(abs)}}(\omega)= I_{xx}^{\mathrm{(abs)}}(\omega)= I_{yy}^{\mathrm{(abs)}}(\omega)$. In Fig. 1(a), because the dipole operator in the response function, $\hat \mu_z$, does not convert the magnetic state, we cannot observe any peak associated with $m$, while the states $m \ne 0$ are thermally well excited. Thus the calculated results of the 3D rotor exhibit discretized rotational bands from the quantum transitions $| j ,m \rangle$ to $| j \pm 1 ,m \rangle$ with the energy difference $ \Delta E_{j\pm 1,m\to j,m} = 2\hbar \omega_0 (j + 1)$, where $\Delta E_{j',m'\to j,m}\equiv E_{j' ,m'} - E_{j,m}$. The rotational bands in the 3D case appear in the even energy states, while those in the 2D case appear in the odd energy states arising from quantum transitions $| j  \rangle$ to $| j \pm 1 \rangle$ with energy differences $\Delta E_{j+1 \to j}  = \hbar \omega_0 (2 j + 1)$.\cite{Iwamoto} Because the QMEs derived from the 2D RISB and 3D RISB Hamiltonian are similar, the calculated spectra exhibit similar behavior in both 2D and 3D RISB cases, in addition to the peak positions. As explained in the 2D case, the peak profile is expressed as a sum of Lorentzian functions.\cite{Iwamoto} 

In Figs. 1(b) and 1(c), in order to observe the magnetic state, we depict rotational absorption spectra of the $z$ and $x$ polarized light under the $z$ polarized Stark field. The energy eigenstates of the rotor system can be estimated using the second-order perturbation theory for the $z$ polarized field as $ E_{0,0} = - E_{S}/3$ and
\begin{align}
E_{j,m} &= j(j+1)\hbar \omega_0 + \frac{j(j+1)- 3m^2}{j(j+1)(2j-1)(2j+3)} E_{S}, 
\label{StarkE}
\end{align}
where $E_{S} \equiv\mu_0^2 E_z^2/2 \hbar \omega_0 $.
This implies that the Stark field resolves a degeneracy of the angular momentum states for different magnetic states.\cite{Townes}
For the  case in Fig. 1(b), the peak positions for small $j$ are evaluated as $\Delta E_{1,0 \to 0,0}^{zz} =2\hbar\omega_0 + \frac{8}{15}E_{S}$ , $\Delta E_{2,1 \to 1,1}^{zz} = 4\hbar\omega_0 -\frac{32}{210} E_{S} $ and $\Delta E_{2,0 \to 1,0}^{zz} =4\hbar\omega_0 + \frac{26}{210} E_{S}$. Thus, we observe two splitting peaks at $\omega= \Delta E_{2,1 \to 1,1}^{zz}$ and $\Delta E_{2,0 \to 1,0}^{zz}$ that arise from the Stark field resolving the degeneracy of the angular momentum states at $\omega = 4\omega_0$, while the single peak at $\omega =\Delta E_{1,0 \to 0,0}^z$ is observed for $j=0$ and $m=0$.
Because the contribution of $m$ for $\Delta E_{j \pm 1,m \to j,m}$ evaluated from Eq.\eqref{StarkE} is larger for small $j$, the splitting peaks of $j=3$ for different $m$ are observed only near $\omega = 6\omega_0$.  {Note that the small peak at $\omega = 6.5\omega_0$ arises from the transition $j = 0 \to 2$.
This transition occurs because the Stark field, $U(\hat{\theta}, \hat{\phi})=E_{z}\hat \mu^z$ involved in Green's function, $\hat{\mathcal{G}}(t)$ induces the angular momentum transition, in addition to the light interactions in Eq. \eqref{eq:R1}.} 
In the case of Fig. 1(c), the peak positions are evaluated as $\Delta E_{1,1 \to 0,0}^{xx} =2\hbar\omega_0 + \frac{7}{30}E_{S}$, $\Delta E_{2,2 \to 1,1}^{xx} =4\hbar\omega_0 + \frac{11}{210} E_{S} $, $\Delta E_{2,1 \to 1,0}^{xx} =4\hbar\omega_0 - \frac{37}{210} E_{S}$, and $\Delta E_{2,0 \to 1,1}^{xx} = 4\hbar\omega_0 + \frac{31}{210} E_{S}$, respectively. Because the $x$ polarized light with $\hat \mu_x$ changes the magnetic quantum number, we observe three Stark splitting peaks near $\omega=4\omega_0$. The peak intensities of three splitting peaks are estimated from the angular eigenstates as $\langle 2,2 | Y^1_1 | 1,1 \rangle : \langle 2,1 | Y^1_1 | 1,0 \rangle : \langle 2,0 | Y^1_1 | 1,1 \rangle = 1 : \sqrt{{1}/{2} } : \sqrt{{1}/{6} }$: The intensity of the third peak is very small. 

In all Figs.1(a)-1(b), peak positions and profiles are similar in the high-frequency region, because the effects of $m$ become small for large $j$ as can be seen from Eq.\eqref{StarkE}.

\subsubsection{Two-dimensional rotational spectrum}
\begin{figure}
\centering
\includegraphics[width=0.99\textwidth]{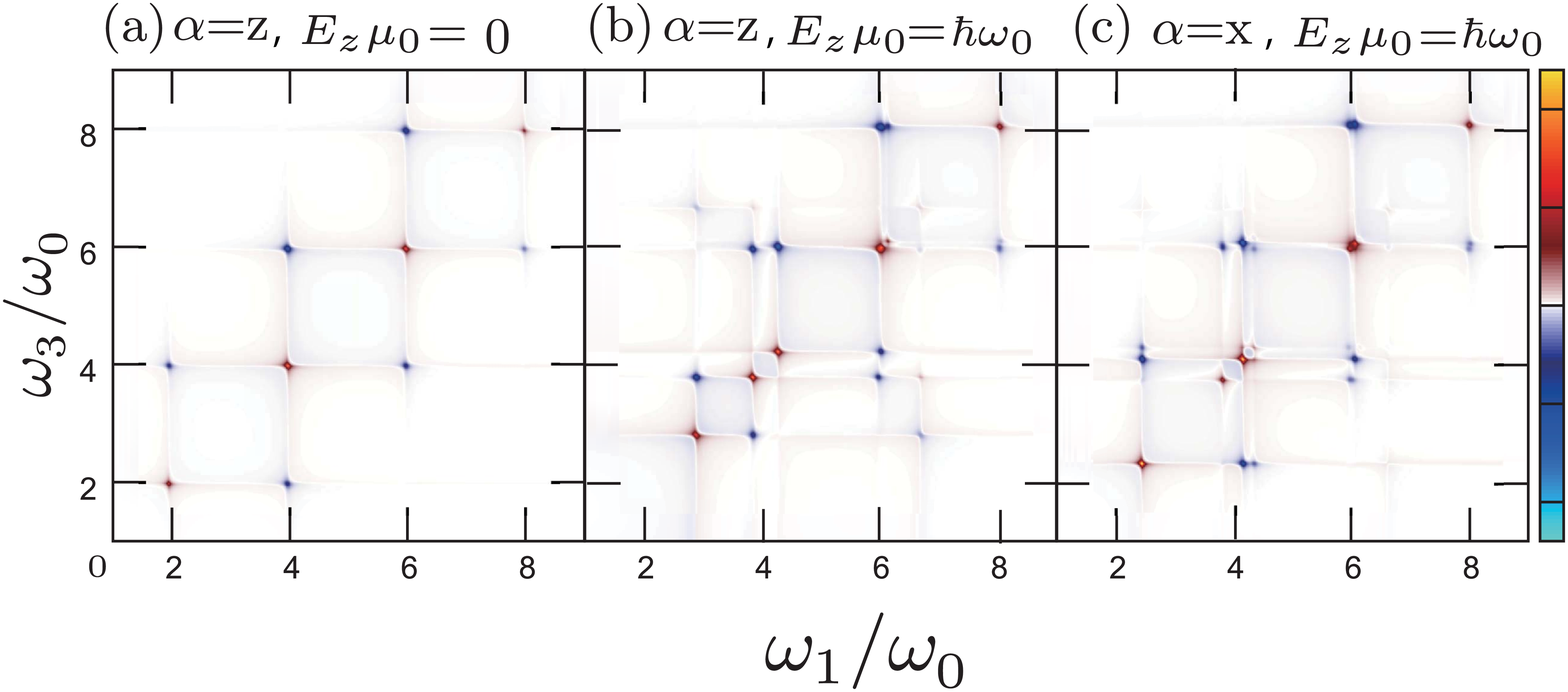}
\caption{
Two-dimensional correlation spectra $I_{\alpha \alpha \alpha \alpha }^{\mathrm{(Corr)}}(\omega _{3},t_{2}=0,\omega _{1})$  for the (a) $\alpha=z$ polarized light without the Stark field, the (b) $\alpha=z$ and (c) $\alpha=x$ polarized light with the Stark field. The blue and red peaks represent the absorption and emission peaks, respectively. The intensity of each peak is normalized with respect to its maximum peak intensity. 
}
\end{figure}

\begin{figure}
\includegraphics[width=0.99\textwidth]{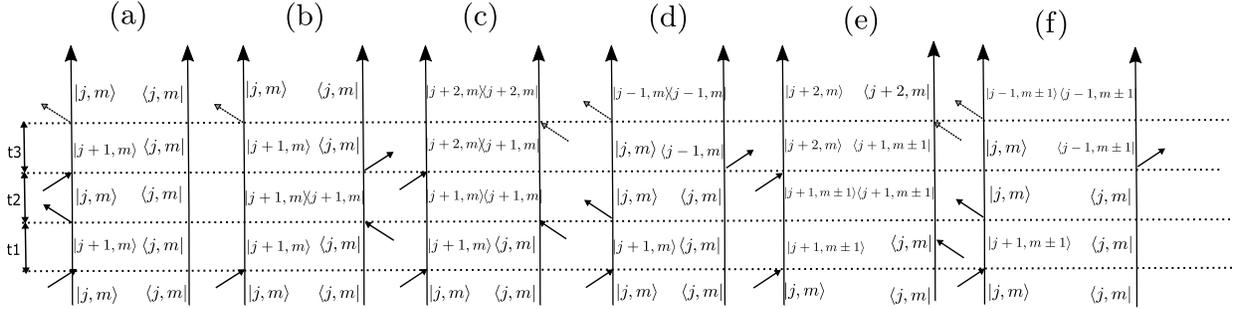}
\caption{
The some of Liouville paths for (a)-(b) the diagonal and (c)-(f) the off-diagonal peaks in 2D spectrum, $I_{\alpha \alpha \alpha \alpha}^{\mathrm{(Corr)}}(\omega _{3},t_{2},\omega _{1})$, where the diagrams (c) and (f) are specifically for the case with the Stark field. The arrows represent optical interactions. In these diagrams, the left line presents the time evolution of the left-hand side wave function (ket), whereas the right one represents the right-hand side (bra). We assume that the system is initially in the population state denoted by $| j ,m \rangle \langle j, m |$. 
}
\end{figure}

In Figs. 2(a)-2(c), we present 2D correlation spectra with and without the Stark field for same sets of parameters in Figs. 1(a)-1(c). The related Liouville path ways are presented in Fig. 3. Here and hereafter we set $t_2=0$.

In Fig. 2(a), we observe the positive diagonal peaks along $\omega_1=\omega_3$ and two negative off-diagonal peaks along $\omega_1=\omega_3 \pm 2\omega_0$. As in the case of Fig. 1(a), we observe the transition of the angular momentum state from $| j ,m \rangle$ to $| j \pm 1 ,m \rangle$ only. The diagonal peaks in Fig. 2(a) arise from the diagram given in Figs.3(a) and 3(b), whereas the upper and lower off-diagonal peaks are from the diagram presented in Fig. 3(c) and 3(d), respectively. Because the $t_3$ period of the diagrams in Figs. 3(a) and 3(b) represent the photon emission process, the diagonal peaks are positive, while the other off-diagonal peaks are negative due to the photon absorption process described by the $t_3$ period of the diagrams in Figs. 3(c) and 3(d). As illustrated in the diagram, the positions of the diagonal and off-diagonal peaks are determined from the coherent states in the $t_1$ and $t_3$ period. (See also 2D spectrum of a 2D rotor case presented in Ref. \onlinecite{Suzuki-2003})

In Figs. 2(b) and 2(c), we depict 2D correlation spectra with the Stark field.
Near $\omega_1=\omega_3 =4\omega_0$, we observe the two Stark splitting peaks in Fig. 2(b), whereas the three Stark splitting peaks in Fig. 2(c), although the third peak in Fig. 2(c) is unnoticeably small. {The location of the diagonal peaks in the figures can be elucidated as the same manner in Figs. 1(b) and 1(c). The varieties of the off-diagonal peaks in Figs. 2(b) and 2(c) represent different transitions that involve both angular momentum and magnetic states most notably the transition  $| j ,m \rangle$ to $| j \pm 1 ,m \pm1 \rangle$, as depicted in Figs. 3(e) and 3(f).
These off-diagonal peaks are characteristic in the 3D rotor case: they appear either parallel to the $\omega_1$ or $\omega_3$ axis, while the diagonal peaks appear along the $\omega_1=\omega_3$ direction. This is because the final state of the Liouville path has to be diagonal as $|j+2 ,m \rangle \langle j+2, m|$ and $|j-1 ,m\pm 1 \rangle \langle j-1, m \pm 1|$  as the diagonal peak case in Fig. 3(c) and 3(d), while the eigenstate in the $t_1$ period involves various magnetic states as $| j + 1 ,m \pm1 \rangle \langle j, m |$ in the off-diagonal peak case.}

As in the case of rotational absorption in Figs. 1(a)-1(c), the peak positions in the high-frequency region are similar in Fig. 2(a)-2(c), while the peaks in the low-frequency region are blue shifted in the Stark case in particular in Fig. 2(b), because the effects of the transition in the $m$ state become important only for small $j$. 

As we demonstrated, the transition pathway involving the angular momentum and magnetic states under the Stark field can be clearly identified by 2D spectroscopy.

\subsection{Effects of system-bath coupling}

\subsubsection{Rotational spectrum}
\begin{figure}
\centering
\includegraphics[width=0.99\textwidth]{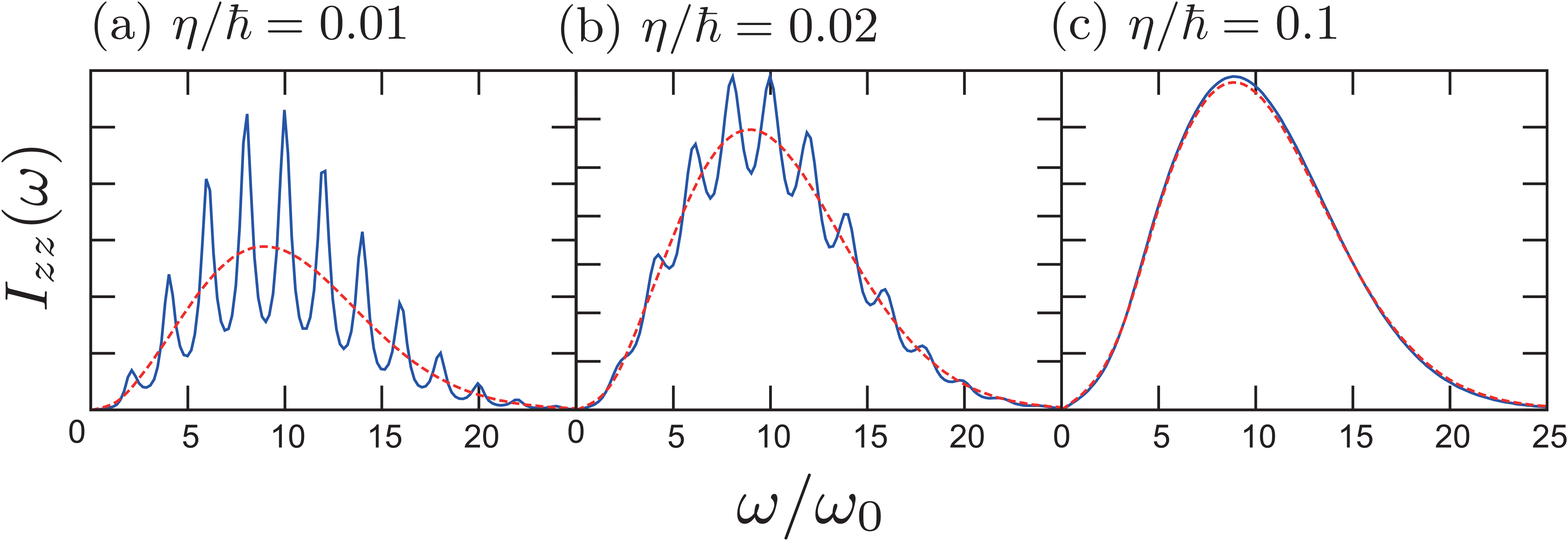}
\caption{
Rotational absorption spectra, $I_{zz}^{\mathrm{(abs)}}(\omega)$, under the (a) weak  ($\eta / \hbar = 0.01$), (b) intermediate ($\eta / \hbar = 0.02$) and (c) strong ($\eta / \hbar = 0.1$) system-bath coupling conditions at the inverse temperature is $\beta\hbar \omega_0 = 0.05$. The blue solid and red dashed curves represent the QME result and the classical ELE result, respectively. The intensity of each line is normalized with respect to its maximum peak intensity.
}
\end{figure}

In Fig. 4, we plot rotational absorption calculated from the QME and classical ELE approaches for the coupling strength, (a) $\eta / \hbar = 0.01$, (b) $\eta / \hbar = 0.02$ and (c) $\eta / \hbar = 0.1 $ in the high temperature case ($\beta\hbar \omega_0 = 0.05$). The spectra in the classical ELE case were calculated from the analytical expression presented in Eq. \eqref{Euler-Langevin}.

In the weak coupling case, depicted in Figs. 4(a) and 4(b), the 3D RISB results exhibit discretized rotational bands arising from quantum transitions, while we observe a broadened peak only in the classical case, because the transition energy of rotational motion is continuous. Even in the quantum case, the rotational peaks are broadened, because the rotational energy levels are mixed and perturbed by the system-bath interactions. 
In the strong coupling case depicted in Fig. 4(c), all of the rotational peaks broaden and merge into a single peak. In such a case, the energy states of the rotor in the quantum case become continuous, because the states of the rotor and bath are entangled due to the strong system-bath interaction. Under the high-temperature approximation, the quantum result approaches the ELE result. 

\subsubsection{Two-dimensional rotational spectrum}
\begin{figure}
\centering
\includegraphics[width=0.99\textwidth]{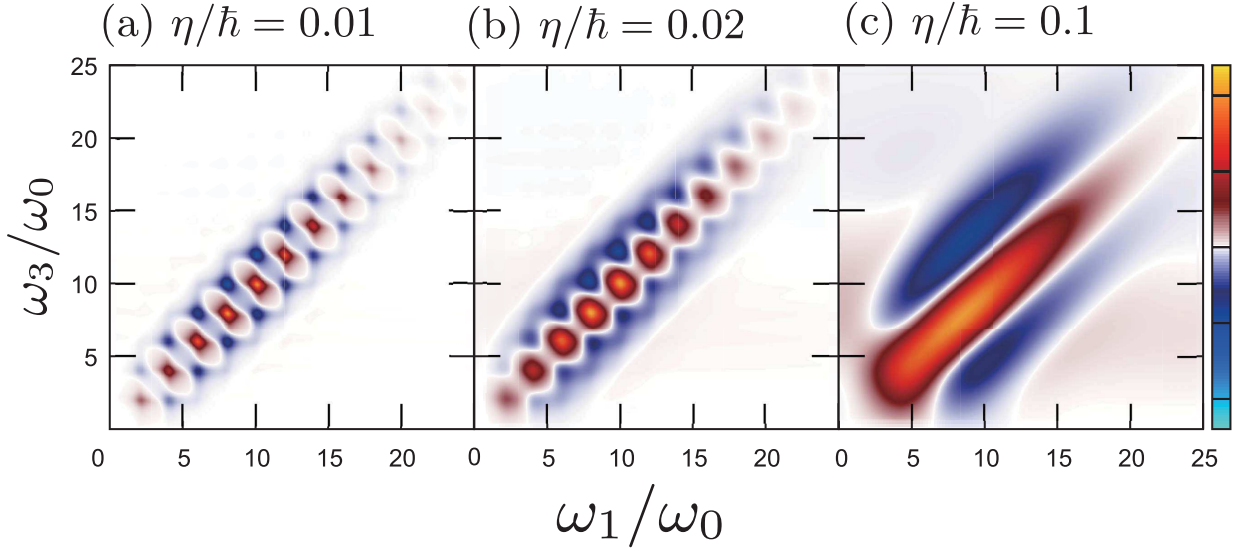}
\caption{
Two-dimensional correlation spectra, $I_{zzzz}^{\mathrm{(Corr)}}(\omega _{3},t_{2}=0,\omega _{1})$, calculated from the QME for (a) $\eta / \hbar = 0.01$, (b) $\eta / \hbar = 0.02$ and (c) $\eta / \hbar = 0.1$ at the inverse temperature is $\beta\hbar \omega_0 = 0.05$. The intensity of each peak is normalized with respect to its maximum peak intensity.
}
\end{figure}

In Figs. 5(a)-5(c), we present the 2D spectra for the same sets of parameters in Fig. 4(a)-(c). In Fig. 5(a), in contrast to the weak coupling case in Fig. 2(a), the profiles of both positive diagonal peaks and negative off-diagonal peaks were changed. The diagonal peaks became star-like shape due to the fast rotational dephasing from Markovian noise that was observed in 2D vibrational spectroscopy,\cite{Ishizaki-2006}  while the off-diagonal peaks become asymmetric shape, because the peaks arise from the magnetic transition illustrated in Figs. 2(b) and 2(c) were merged. In the present case, however, instead of the Stark field, the dipole operator in the system-bath coupling in the three direction, $\hat V_{\alpha}= \hat \mu_{\alpha}/\mu_0$ for $\alpha=x, y$ and $z$, created such transitions in a very complex manner. 

In Fig. 5(b), as we increase the coupling strength, the off-diagonal peaks elongated to the $\omega_1=\omega_3$ direction, while the profiles of diagonal peaks become circular. The very characteristic feature of these spectra is white square regions in the vicinity of the diagonal peaks. Such regions arise because the transition through the magnetic quantum number appear in the direction parallel to either the $\omega_1$ or $\omega_3$ axis, as depicted in Figs. 2(b) and 2(c). Although the shape is not square, the white regions are also observed in Fig. 5(a): The asymmetric profiles of the negative peaks can be the evidence of the transitions through the magnetic states that arise from the system-bath interactions. This result demonstrate that 2D rotational spectroscopy has a capability to identify the bath induced magnetic transitions, while rotational absorption on the 3D rotor case is similar to that of the 2D rotor case, in which the magnetic states cannot play a role. 

In Fig. 5(c), when the system-bath coupling becomes very strong, the positive and negative peaks are broadened and merged, respectively. In comparison to cases in Figs. 5(a) and 5(b), the widths of the blue peaks become larger. This is also regarded as the evidence of the magnetic transitions. As explained above, the system-bath coupling plays the same role as the Stark field: The widths become larger, because
the stark splitting amplitude of the magnetic transition becomes large in this strong coupling regime. We note that, for this overdamped case, information concerning the energy gap between the rotational levels cannot be obtained in the rotational absorption: These spectra reveal a continuous broad peak that reflects the initial thermal distribution. Contrastingly, although we cannot explore the details of the transition states, we observe the discretization of the energy states as the three elongated broadened peak in the $\omega_1=\omega_3$ direction in the 2D rotational spectrum.\cite{Suzuki-2003}

\section{Conclusion}
In this paper, we introduced a rotationally invariant system-bath model in three-dimensional space to describe the dynamics of a linear rigid rotor in a dissipative environment. While quantum effects vanish in the high-temperature regime for a system described by a Brownian model, the present rotor model exhibits quantum features even at very high temperature. This is because the quantum coherence of the rotor system is long-lived, due to the inertial property of the rotational motion, which is confined only by the symmetry of the space. 
In order to characterize the dissipative dynamics of the 3D rotor system, we calculated linear absorption and 2D rotational spectra using the high-temperature limit of the QME formalism. 

In the very weak system-bath coupling case, we analyzed the role of magnetic states by applying a polarized Stark field. In the linear spectra, while we observe transition peaks of the angular momentum only in the case without the external field, we can distinguish various magnetic states as separated peaks by applying the Stark field to resolve the degeneracy of the angular momentum states. The calculated 2D spectra utilized with the Stark field made it possible to identify the transition states through the change of both angular momentum states and magnetic states,

We then studied the effects of the system-bath coupling strength. In linear spectra, a transition of the peak profiles from discretized rotational bands to a Lorentzian-like peak through a Gaussian-like peak is observed as a function of the system-bath coupling in a unified manner. When the system-bath coupling becomes strong, the bath interactions in the $x$, $y$ and $z$ directions convert the varieties of magnetic states in a very complex manner. The contribution of the magnetic transitions is observed as the surrounding spectra of the white square regions in the vicinity of the diagonal peaks in the underdamped regime, while that contribution is observed as the two elongated broadened peaks parallel to the diagonal peak in the overdamped regime. It was shown that even in the high-temperature overdamped case, in which the classical and quantum linear spectra coincide, we find clear evidence of the quantum transition as the off-diagonal elements of the 2D peaks. While this feature was discovered in a 2D rotor system described by the RCL model,\cite{Suzuki-2003} here we demonstrated it in a more realistic situation using the RISB model. 

In this paper, we limited our analysis to the high-temperature Markovian case, employing a linear rigid rotor system. The extension of the present model to symmetric and asymmetric top models is necessary to investigate realistic molecules, for example, to study the rotational dynamics of molecules embedded in anisotropic crystals. However, to treat symmetric tops or asymmetric tops, we need to look for the model Hamiltonian from which we can derive classical equation such as Langevin equation.
Because the eigenstate representation of such a system is extremely complicated, it may be easier to employ a coordinate space representation, while the finite difference representation of the momentum and potential operator must be carefully constructed. Such a description would also be suitable for inclusion of electronic excitation states\cite{Ikeda2019} and to treat time-dependent anisotropic potentials, which is important to study laser-induced molecular alignment.\cite{Fleischer2017,Seideman2017}  

Although the computations become numerically intensive, the present theory provides a framework for studying the non-Markovian effects of both isotropically and anisotropically correlated noise. Moreover, if necessary, we can include low-temperature correction terms in the HEOM expression to study very low-temperature systems,\cite{Tanimura-1989, Tanimura-1990,Ishizaki-2005,Tanimura-2006,Tanimura-2014,Tanimura-2015,Lipeng2019} for example, to study methane molecules embedded in parahydrogen crystals.\cite{momose}  The inclusion of the effect of intramolecular vibrational motion is also important to understand complex dynamics through chemical reaction processes. We may have to employ a different formalism based on the HEOM, however, in order to reduce the numerical cost\cite{Nakamura} and in order to treat more complex systems, for example, a rotor system coupled with a spin bath.\cite{Cao2018A,Cao2018B} 
Such investigations employing the framework of the present work will be carried out in the future. 

\section*{Acknowledgments}
Y.~T.~is supported by JSPS KAKENHI Grant Number A26248005.

\appendix
\section{Euler-Langevin Equation}
For the Markovian and isotropic case described by an Ohmic spectral distribution $ J^{\alpha}(\omega)= \eta \omega$, where $\eta$ is the friction coefficient, the equations of motion are expressed as\cite{Lindenberg}
\begin{align}
I \dot{\omega_x} - I \omega_y^2 \cot \theta + \eta \omega_x = \lambda_x, \notag \\
I \dot{\omega_y} + I \omega_x \omega_y \cot \theta + \eta \omega_y = \lambda_y, \\
\bf{\omega} = (\dot{\theta} , \dot{\phi} \sin \theta , \dot{\phi} \cos \theta). \notag
\end{align}
where the random torques have the properties
\begin{align}
\langle \lambda_j \rangle = 0 , 
\end{align}
and
\begin{align}
\langle \lambda_j \lambda_m \rangle = 2 k_B T \eta \delta_{jm} \delta(t-t') 
\end{align}
for $ j,m = x,y$.

In the classical case, the response function from the correlation function can be defined as 
\begin{align}
C_{\alpha' \alpha }(t) &\equiv  \langle \mu_{\alpha}(0) \mu_{\alpha'}(t) \rangle_{cl},
\end{align}
where $\langle  \rangle_{cl}$ represents the thermal average over the classical distribution. The rotational absorption spectrum is then expressed as
$I_{\alpha \alpha}^{(abs)}(\omega)= {i \omega \mu_0^2 }C_{\alpha \alpha }[\omega]/ {\beta}$, where  $C_{\alpha \alpha } [\omega]$ is the Fourier transforms $C_{\alpha \alpha }(t)$. The analytical expression for the response functions has been obtained in the Ohmic case.\cite{Coffey} 
The rotational absorption spectrum for the isotropic environment is then obtained by
\begin{align}
I_{zz}^{(abs)}(\omega)= \frac{ \omega \mu_0^2 }{\beta} \mathrm{Re} \frac{\alpha'}{i \omega \alpha' + \frac{1}{\beta' + i \omega \alpha' + \frac{1}{2\beta' + i \omega \alpha' + \frac{2}{3\beta' + i \omega \alpha' + \dots } } }},
\label{Euler-Langevin}
\end{align}
where $\alpha' = {\sqrt{\beta \hbar \omega_0}}/{2}$ and $\beta' = {\eta \alpha'}/{I}$.

\section{Eigenstate Representation of QME}
We employ the spherical harmonics, $ Y^{m}_{j}(\theta ,\phi)$ and the formula expressed as\cite{Zare}
\begin{align}
&\int^{2 \pi}_0 d \phi \int^{\pi}_0 d \theta \sin \theta Y^{m_3}_{j_3}(\theta ,\phi) Y^{m_2}_{j_2}(\theta ,\phi) Y^{m_1}_{j_1}(\theta ,\phi) \notag \\
&= \left[ \frac{(2 j_1 + 1)(2 j_2 + 1)(2 j_3 + 1)}{4 \pi} \right]^{\frac{1}{2}} \begin{pmatrix} j_1 \ j_2 \ j_3 \\ m_1 \ m_2 \ m_3\end{pmatrix}
\begin{pmatrix} j_1 \ j_2 \ j_3 \\ 0 \ 0 \ 0\end{pmatrix}
\end{align}
where $\begin{pmatrix} j_1 \ j_2 \ j_3 \\ m_1 \ m_2 \ m_3\end{pmatrix}$ is Wigner 3-j symbols.  
Then the QME for the element $\rho^{j_1,j_2}_{m_1,m_2}(t) \equiv \langle j_1 m_1 |\rho| j_2 m_2 \rangle$ is expressed as
\begin{align}
\frac{\partial}{\partial t} \rho^{j_1,j_2}_{m_1,m_2}(t)  &= \left[-i \omega_0 ( j_1(j_1+1) - j_2(j_2+1) ) - 2\eta \left(\frac{1}{\beta} - 1 \right) \right]\rho^{j_1,j_2}_{m_1,m_2}(t) \notag \\
& + \eta \left(\frac{2}{\beta} - j_1 - j_2 \right) C_{mm}^{j_1,m_1} C_{mm}^{j_2,m_2} \rho^{j_1-1,j_2-1}_{m_1-1,m_2-1} \notag \\
&- \eta \left(\frac{2}{\beta} - j_1 + j_2 + 1 \right) C_{mm}^{j_1,m_1} C_{pm}^{j_2,m_2} \rho^{j_1-1,j_2+1}_{m_1-1,m_2-1} \notag \\
&+ \eta \left(\frac{2}{\beta} - j_1 - j_2 \right) C_{m0}^{j_1,m_1} C_{m0}^{j_2,m_2} \rho^{j_1-1,j_2-1}_{m_1,m_2} \notag \\
&+ \eta \left(\frac{2}{\beta} - j_1 + j_2 + 1 \right) C_{m0}^{j_1,m_1} C_{p0}^{j_2,m_2} \rho^{j_1-1,j_2+1}_{m_1,m_2} \notag \\
& + \eta \left(\frac{2}{\beta} - j_1 - j_2 \right) C_{mp}^{j_1,m_1} C_{mp}^{j_2,m_2} \rho^{j_1-1,j_2-1}_{m_1+1,m_2+1} \notag \\
& - \eta \left(\frac{2}{\beta} - j_1 + j_2 + 1 \right) C_{mp}^{j_1,m_1} C_{pp}^{j_2,m_2} \rho^{j_1-1,j_2+1}_{m_1+1,m_2+1} \notag \\
& - \eta \left(\frac{2}{\beta} + j_1 - j_2 + 1 \right) C_{pm}^{j_1,m_1} C_{mm}^{j_2,m_2} \rho^{j_1+1,j_2-1}_{m_1-1,m_2-1} \notag \\
& + \eta \left(\frac{2}{\beta} + j_1 + j_2 + 2 \right) C_{pm}^{j_1,m_1} C_{pm}^{j_2,m_2} \rho^{j_1+1,j_2+1}_{m_1-1,m_2-1} \notag \\
&+ \eta \left(\frac{2}{\beta} + j_1 - j_2 + 1 \right) C_{p0}^{j_1,m_1} C_{m0}^{j_2,m_2} \rho^{j_1+1,j_2-1}_{m_1,m_2} \notag \\
&+ \eta \left(\frac{2}{\beta} + j_1 + j_2 + 2 \right) C_{p0}^{j_1,m_1} C_{p0}^{j_2,m_2} \rho^{j_1+1,j_2+1}_{m_1,m_2} \notag \\
& - \eta \left(\frac{2}{\beta} + j_1 - j_2 + 1 \right) C_{pp}^{j_1,m_1} C_{mp}^{j_2,m_2} \rho^{j_1+1,j_2-1}_{m_1+1,m_2+1} \notag \\
& + \eta \left(\frac{2}{\beta} + j_1 + j_2 + 2 \right) C_{pp}^{j_1,m_1} C_{pp}^{j_2,m_2} \rho^{j_1+1,j_2+1}_{m_1+1,m_2+1} 
\label{QMEANG}
\end{align}
where
\begin{align}
C^{j,m}_{pm} &= \sqrt{\frac{(j-m+1)(j-m+2)}{2(2j+1)(2j+3)} } \notag \\
C^{j,m}_{p0} &= \sqrt{\frac{(j+m+1)(j-m+1)}{(2j+1)(2j+3)} } \notag \\
C^{j,m}_{pp} &= \sqrt{\frac{(j+m+1)(j+m+2)}{2(2j+1)(2j+3)} } \notag \\
C^{j,m}_{mm} &= \sqrt{\frac{(j+m-1)(j+m)}{2(2j-1)(2j+1)} } \notag \\
C^{j,m}_{m0} &= \sqrt{\frac{(j+m)(j-m)}{(2j-1)(2j+1)} } \notag \\
C^{j,m}_{mp} &= \sqrt{\frac{(j-m-1)(j-m)}{2(2j-1)(2j+1)} }.  
\end{align}
As the right-hand side terms in Eq.\eqref{QMEANG} indicate, the heat baths allow the energy transitions only from $(j_1, j_2; m_1,m_2)$ to $(j_1\pm 1, j_2\pm 1; m_1\pm 1,m_2\pm 1)$. In particular, the magnetic quantum states $(m_1,m_2)$ are coupled only to $(m_1,m_2)$ and $(m_1 \pm 1,m_2 \pm 1)$. Such a feature is the same as the 2D rotor case described by the QME.\cite{Iwamoto}

\bibliography{aipsamp}

\end{document}